\documentclass[%
 reprint,
superscriptaddress,
 amsmath,amssymb,
 aps,
prl,
twocolumn
]{revtex4}

\usepackage{graphicx}% Include figure files
\usepackage{dcolumn}% Align table columns on decimal point
\usepackage{bm}% bold math
\usepackage[version=4]{mhchem}
\usepackage{comment}
\usepackage{color,soul}
\usepackage{gensymb}

\begin{document}

%\preprint{APS/123-QED}

\title{Influence of Polymorphism on the Electronic Structure of \ce{Ga2O3}}%

\author{Jack E. N. Swallow}
 \affiliation{Stephenson Institute for Renewable Energy and Department of Physics, University of Liverpool, Liverpool L69 7ZF, United Kingdom
}%Lines break automatically or can be forced with \\

\author{Christian Vorwerk}%
\affiliation{%
Institut f\"{u}r Physik and IRIS Adlershof, Humboldt-Universit\"{a}t zu Berlin, Berlin, Germany
}%

\author{Piero Mazzolini}%
\affiliation{%
Paul-Drude-Institut fur Festk\"{o}rperelektronik, Leibniz-Institut im Forschungsverbund Berlin e.V., Hausvogteiplatz 5-7, 10117 Berlin, Germany
}%

\author{Patrick Vogt}%
\affiliation{%
Paul-Drude-Institut fur Festk\"{o}rperelektronik, Leibniz-Institut im Forschungsverbund Berlin e.V., Hausvogteiplatz 5-7, 10117 Berlin, Germany
}%

\author{Oliver Bierwagen}%
\affiliation{%
Paul-Drude-Institut fur Festk\"{o}rperelektronik, Leibniz-Institut im Forschungsverbund Berlin e.V., Hausvogteiplatz 5-7, 10117 Berlin, Germany
}%

\author{Alexander Karg}%
\affiliation{%
Institut für Festk\"orperphysik, Universit\"at Bremen, Otto-Hahn-Allee 1, 28359 Bremen, Germany
}%

\author{Martin Eickhoff}%
\affiliation{%
Institut für Festk\"orperphysik, Universit\"at Bremen, Otto-Hahn-Allee 1, 28359 Bremen, Germany
}%

\author{J\"org Sch\"ormann}%
\affiliation{%
Institute of Experimental Physics I, Justus Liebig University Giessen, 35392 Giessen, Germany
}%

\author{Markus R. Wagner}%
\affiliation{%
Technische Universit\"{a}t Berlin, Institute of Solid State Physics, Hardenbergstr. 36, 10623 Berlin, Germany
}%

\author{Joseph W. Roberts}%
\affiliation{%
School of Engineering, University of Liverpool, Brownlow Hill, Liverpool L69 3GH, United Kingdom
}%

\author{Paul R. Chalker}%
\affiliation{%
School of Engineering, University of Liverpool, Brownlow Hill, Liverpool L69 3GH, United Kingdom
}%

\author{Matthew J. Smiles}%
\affiliation{%
Stephenson Institute for Renewable Energy and Department of Physics, University of Liverpool, Liverpool L69 7ZF, United Kingdom
}%

\author{Philip Murgatroyd}%
\affiliation{%
Stephenson Institute for Renewable Energy and Department of Physics, University of Liverpool, Liverpool L69 7ZF, United Kingdom
}%

\author{Sara A. Razek}%
\affiliation{%
Department of Physics, Binghamton University, State University of New York, Binghamton, New York 13850, USA
}%
\affiliation{%
Materials Science and Engineering, Binghamton University, State University of New York, Binghamton, New York 13850, USA
}%

\author{Zachary W. Lebens-Higgins}%
\affiliation{%
Department of Physics, Binghamton University, State University of New York, Binghamton, New York 13850, USA
}%
\affiliation{%
Materials Science and Engineering, Binghamton University, State University of New York, Binghamton, New York 13850, USA
}%

\author{Louis F. J. Piper}%
\affiliation{%
Department of Physics, Binghamton University, State University of New York, Binghamton, New York 13850, USA
}%
\affiliation{%
Materials Science and Engineering, Binghamton University, State University of New York, Binghamton, New York 13850, USA
}%

\author{Leanne A. H. Jones}%
\affiliation{%
Stephenson Institute for Renewable Energy and Department of Physics, University of Liverpool, Liverpool L69 7ZF, United Kingdom
}%

\author{Pardeep K. Thakur}%
\affiliation{%
Diamond Light Source Ltd., Diamond House, Harwell Science and Innovation Campus, Didcot OX11 0DE, United Kingdom
}%

\author{Tien-Lin Lee}%
\affiliation{%
Diamond Light Source Ltd., Harwell Science and Innovation Campus, Didcot OX11 0DE, United Kingdom
}%

\author{Joel B. Varley}%
\affiliation{%
Lawrence Livermore National Laboratory, Livermore, CA, 94550, United States of America
}%

\author{J\"{u}rgen Furthm\"{u}ller}%
\affiliation{%
Institut f\"{u}r Festk\"{o}rpertheorie und -optik and European Theoretical Spectroscopy Facility, Friedrich-Schiller-Universit\"{a}t Jena, Max-Wien-Platz 1, D-07743 Jena, Germany
}%

\author{Claudia Draxl}%
\affiliation{%
Institut f\"{u}r Physik and IRIS Adlershof, Humboldt-Universit\"{a}t zu Berlin, Berlin, Germany
}%

\author{Tim D. Veal}%
\affiliation{%
Stephenson Institute for Renewable Energy and Department of Physics, University of Liverpool, Liverpool L69 7ZF, United Kingdom
}%

\author{Anna Regoutz}%
 \email{a.regoutz@ucl.ac.uk}
\affiliation{%
Department of Chemistry, University College London, 20 Gordon Street, London WC1H 0AJ, United Kingdom
}%

\date{\today}% 

%############################################## max 150 words, 171 now
\begin{abstract}

ABSTRACT: The search for new wide band gap materials is intensifying to satisfy the need for more advanced and energy efficient power electronic devices. \ce{Ga2O3} has emerged as an alternative to SiC and GaN, sparking a renewed interest in its fundamental properties beyond the main $\beta$-phase. Here, three polymorphs of \ce{Ga2O3}, $\alpha$, $\beta$ and $\varepsilon$, are investigated using X-ray diffraction, X-ray photoelectron and absorption spectroscopy, and \textit{ab initio} theoretical approaches to gain insights into their structure - electronic structure relationships. Valence and conduction electronic structure as well as semi-core and core states are probed, providing a complete picture of the influence of local coordination environments on the electronic structure. State-of-the-art electronic structure theory, including all-electron density functional theory and many-body perturbation theory, provide detailed understanding of the spectroscopic results. The calculated spectra provide very accurate descriptions of all experimental spectra and additionally illuminate the origin of observed spectral features. This work provides a strong basis for the exploration of the \ce{Ga2O3} polymorphs as materials at the heart of future electronic device generations.

\end{abstract}
%##############################################

\maketitle

%############################################## 500 words max for intro (3k max in total, 4-6 figures) 654 now
\section{Introduction}
%##############################################
\ce{Ga2O3} is an ultra-wide band gap, transparent, semiconducting oxide material. Its band gaps of 4.5 - 5.3 eV depending on its crystal structure are significantly larger than those of SiC (3.3 eV) and GaN (3.4 eV) and make it an exciting material for future power electronic applications, as well as for solar blind UV photodetectors and gas sensing devices \cite{KimJMCC2017,HigashiwakiPSSA2014,KokubunAPL2007,OGITAASS2001,MazeinaLang2010}. Its most stable form is the monoclinic $\beta$-phase, which combines the large band gap with high conductivity \cite{UedaAPL1997} and breakdown field \cite{HigashiwakiSST2016}, as well as good transparency in the UV spectral region \cite{UedaAPL1997b}.\\

While the $\beta$-phase has been the focus of almost all recent studies on \ce{Ga2O3}, a range of polymorphs exists, similar to other polymorphic oxide systems including \ce{Al2O3} \cite{LevinJACS1998,FilatoveJPCC2015}, \ce{In2O3} \cite{FuchsPRB2008,KingPRB2009}, and \ce{Sb2O3} \cite{AllenJPCC2013}. Five different \ce{Ga2O3} polymorphs have been reported in the literature \cite{RoyJACS1952,YusaPRB2008,Zhang20}. These include monoclinic $\beta$-\ce{Ga2O3} (space group C2/m), rhombohedral $\alpha$-\ce{Ga2O3} (space group R\={3}c), cubic $\gamma$-\ce{Ga2O3} (space group Fd\={3}), orthorhombic $\varepsilon$-\ce{Ga2O3} (space group Pna2\textsubscript{1}), and the cubic $\delta$-\ce{Ga2O3} (space group Ia\={3}) phase, although it has been previously argued that the $\delta$-phase is formed from a mixture of the $\beta$- and $\varepsilon$-phases \cite{playford2013}. It is also important to note that the $\varepsilon$-\ce{Ga2O3} phase may also refer to the hexagonal (P6\textsubscript{3}mc) structure, which is constructed by twinning of three rotational domains of the orthorhombic $\varepsilon$-phase. Finally, to add to the complexity of different reported phases, the orthorhombic phase is sometimes referred to as the $\kappa$-\ce{Ga2O3} phase \cite{CoraCEC2017, YoshiokaJP2007}, and the two designations used interchangeably.\\

Although interest in \ce{Ga2O3} as an ultra-wide band gap material is great, the polymorphs beyond $\beta$-\ce{Ga2O3} are relatively unexplored, especially experimentally, with very limited information regarding their electronic properties available. This is due in part to limitations in producing high quality samples, but challenges are not limited solely to experimental efforts. The large number of structural phases and the strong anisotropy of $\beta$-\ce{Ga2O3} present difficulties for theoretical approaches. Nevertheless, previous theoretical work carried out on \ce{Ga2O3} polymorphs focuses on various properties, e.g. optical response, electron transport, electronic structure, and X-ray absorption spectra, calculated at different levels of computational expense and accuracy.\cite{FurtmullerPRB2016,YoshiokaJP2007,HaiyingPRB2006,LITIMEINJAC2009,Lyons2019JSSST,CocchiPRB2016,Kang_2017,ghosh_singisetti_2017} Experimental studies predominantly consider material growth and optical properties with more extensive information available for the $\beta$-phase. \cite{YaoECS_2017,CoraCEC2017,NISHINAKA201728,SeguraPRM2017,SwallowAPL2019,HOU200712,CollinsJPCB2006,GOTTSCHALCHJCG2019,Ma2016,Golz2019} Very few studies to date combine theory and experiment, thus leaving some doubt over the accuracy of calculations and uncertainty in the understanding and interpretation of experimental
data.\cite{LingAPL2015,MulazziAPL2019,SwallowAPL2019,CocchiPRB2016,Navarroprb2015} Overall, little information is available for the polymorphs of \ce{Ga2O3} and studies considering a set or subset of the polymorphs are extremely rare. However, comparative studies of the polymorphs are invaluable for discerning trends between the different structures and advancing our understanding of the material.\\

The present work provides an in-depth study of the electronic structure of three \ce{Ga2O3} polymorphs: $\alpha$-\ce{Ga2O3}, $\beta$-\ce{Ga2O3}, and $\varepsilon$-\ce{Ga2O3}. In order to achieve the best experimental results possible, the highest quality available samples were selected for this study including bulk single crystals and epitaxial films grown using molecular beam epitaxy (MBE) and atomic layer deposition (ALD). X-ray diffraction (XRD) is used to provide information on the crystallinity of the samples. High-resolution soft and hard X-ray photoelectron spectroscopy (SXPS at 1.49 keV and HAXPES at 5.93 keV) are reported in conjunction with X-ray absorption spectroscopy (XAS) to probe the electronic structure of both occupied and unoccupied states. Beyond the valence states, semi-core and core state spectra are explored and clear effects of local coordination geometry are found. The combination of SXPS and HAXPES enables the study of both surface and bulk electronic structure and to identify features intrinsic to bulk \ce{Ga2O3}. All spectral data are directly compared to results from ab-initio electronic structure calculations using all-electron DFT and many-body perturbation theory (MBPT) in the GW approach, as well as absorption spectra using MBPT in the Bethe-Salpeter equation (BSE) approach. Ultimately, this work presents a systematic and comprehensive comparison of the electronic structure of the \ce{Ga2O3} polymorphs, providing an insight into electronic trends and their relationship to crystal structure. It lays the foundation for further exploration of \ce{Ga2O3} in applications beyond its $\beta$-phase.\\

%############################################## 360 words
\section{Structure}
%##############################################

\begin{figure*}[ht!]
\centering
\includegraphics[width=0.98\textwidth]{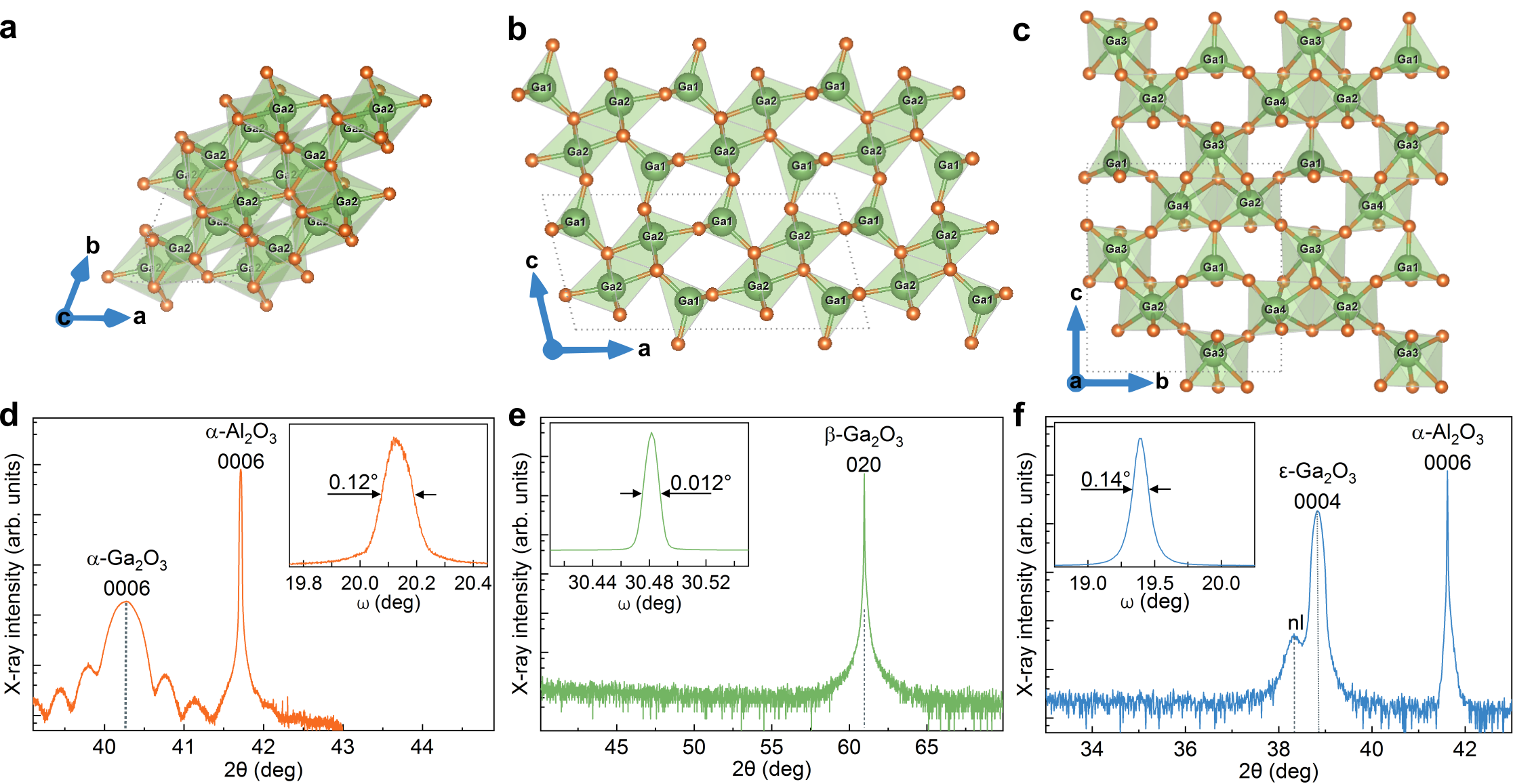}
\caption{\textbf{Structure of the \ce{Ga2O3} polymorphs. a-c,} Schematic representations of the crystal lattice structures of \textbf{a,} $\alpha$, \textbf{b,} $\beta$, and \textbf{c,} $\varepsilon$-polymorphs of \ce{Ga2O3}. Inequivalent Ga positions are numbered with tetrahedral (1) and octahedral coordination (2-4) in the and crystal structures prepared using VESTA \cite{MommaJAC2011}. \textbf{d-f}, X-ray diffraction patterns of \textbf{d,} $\alpha$, \textbf{e,} $\beta$, and \textbf{f,} $\varepsilon$-polymorphs of \ce{Ga2O3} showing symmetric out-of-plane 2$\theta$-$\omega$ scans on a logarithmic scale. The insets show the corresponding $\omega$-rocking curves of the \ce{Ga2O3} peaks on a linear scale along with their full widths at half maximum.}
\label{structure}
\end{figure*}

The crystal structures of the three polymorphs investigated are shown in Fig.~\ref{structure}a-c. The polymorphs are constructed from varying Ga atomic arrangements, i.e. different stacking of the tetrahedrally and octahedrally coordinated building blocks, which contribute to the phase stability as well as the opto-electronic properties of the polymorphs \cite{GellerJCP1960,AhmanACC1996}. The $\beta$-phase crystallises in a monoclinic structure with two inequivalent Ga\textsuperscript{3+} sites in a ratio of 1:1, one in a distorted tetrahedron and the other in a highly distorted octahedron. In rhombohedral corundum $\alpha$-\ce{Ga2O3} the Ga\textsuperscript{3+} ion coordination is octahedral only, and the unit cell has the smallest volume out of the polymorphs discussed in this study. Finally, the orthorhombic $\varepsilon$-\ce{Ga2O3} structure contains both octahedrally and tetrahedrally coordinated Ga\textsuperscript{3+} sites, with three times the number of octahedra compared to tetrahedra in the unit cell, making $\varepsilon$-\ce{Ga2O3} a structural intermediate of the $\alpha$ and $\beta$-phases.\\

XRD patterns were collected to confirm the different structures, phase purity, and crystallinities of the samples and are shown in Fig.~\ref{structure}d-f (wide angle 2$\theta$-$\omega$ scans are included in Fig.~S1 in the Supplementary Information). The $\alpha$-\ce{Ga2O3} phase, shown in Fig.~\ref{structure}d, displays a (0001) film orientation with a single 0006 film peak at 2$\theta$=40.2$^{\circ}$. Fringes in the diffraction pattern can be interpreted as indication of high film quality with good uniformity in the thickness and coverage of the film \cite{ROBERTS2019125254}. The diffraction pattern from bulk $\beta$-\ce{Ga2O3} (010), shown in Fig.~\ref{structure}e, exhibits the allowed 020 peak with an extremely sharp rocking curve width of 0.012$^\circ$, in agreement with the high structural quality of melt grown, single crystal samples. The diffraction pattern from MBE-grown $\varepsilon$-\ce{Ga2O3} shown in Fig.~\ref{structure}f, can be seen to have a strong peak associated with the (0004) orientation at 2$\theta$=38.7$^{\circ}$. A shoulder, labeled ``nl", on the low angle side of this peak is associated with the  $\bar{4}$02 reflection of the $\beta$-\ce{Ga2O3}($\bar{2}$01) nucleation layer (nl) \cite{Vogt2017} or interfacial layer \cite{KrachtPRA2017}. The $\alpha$ and $\varepsilon$ films exhibit $\omega$-rocking curve widths that are typical for heteroepitaxial, single crystalline films on non-lattice matched substrates. The in-plane lattice mismatch between the sapphire substrate and the $\alpha$ or $\varepsilon$ films is 4.7~$\%$\cite{Cheng2017} and 4.1~$\%$,\cite{Mezzadri2016} respectively. The 0006 peak of the c-plane \ce{Al2O3} substrate used for the epitaxy of these phases can be seen at 2$\theta$=41.7$^{\circ}$.\\

\begin{figure*}[ht!]
\centering
\includegraphics[width=0.98\textwidth]{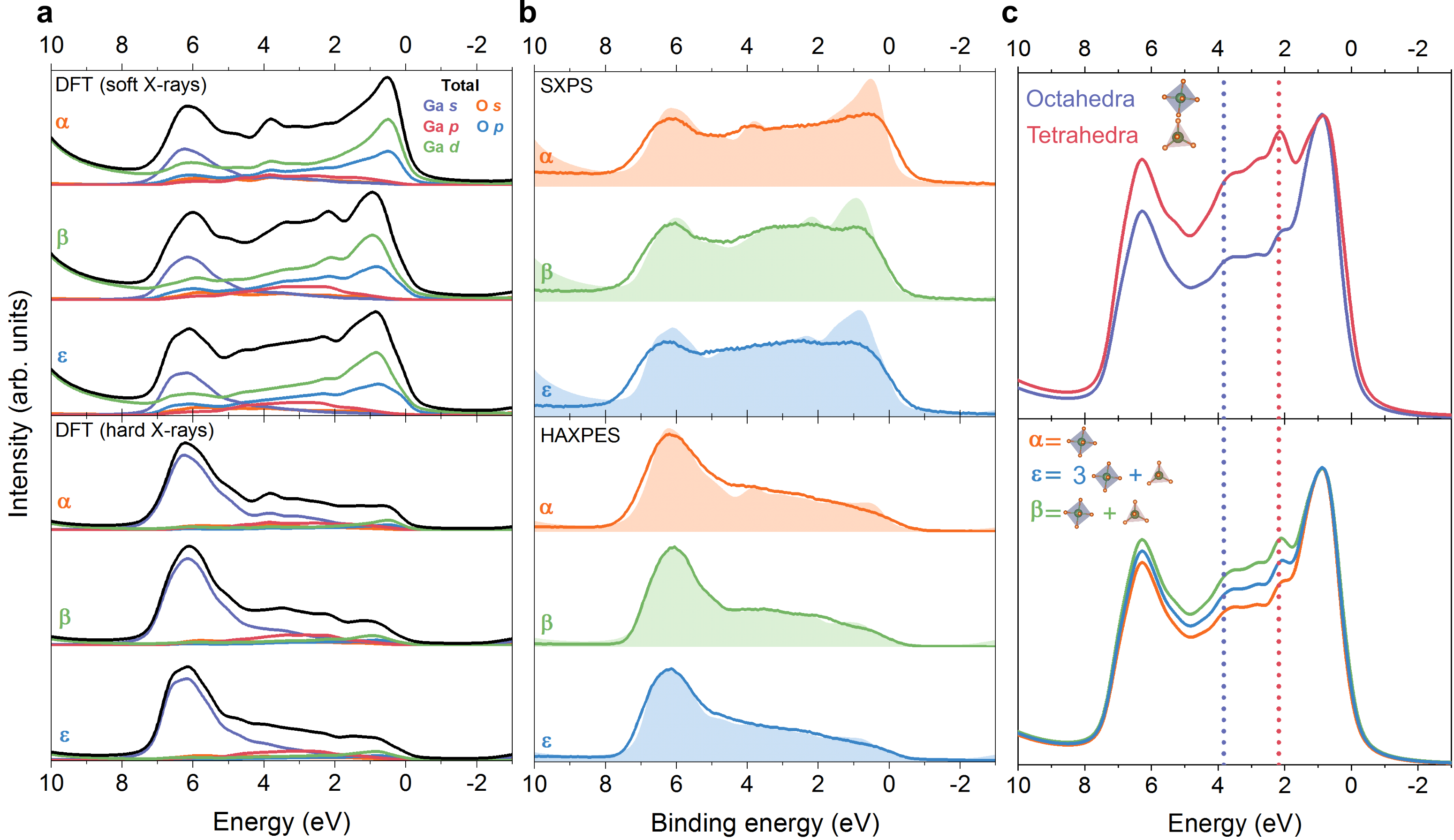}
\caption{\textbf{Valence band structure of the \ce{Ga2O3} polymorphs. a,} Calculated pDOS (coloured) and tDOS (black) weighted for the photoionisation cross sections of soft and hard X-ray photoelectron spectroscopy measurements. \textbf{b,} Experimental SXPS and HAXPES spectra (lines) each aligned using the calculated tDOS (filled curves) as a reference with a rigid shift of the data so that the VBM is at 0 eV (see Supplementary Information). The tDOS is overlayed to aid comparison. \textbf{c,} DOS for octahedrally and tetrahedrally coordinated Ga calculated from $\beta$-\ce{Ga2O3} and combinatorial calculated spectra for all three polymorphs.}
\label{softhardcomp}
\end{figure*}

%############################################## 678 words
\section{Occupied States}
%##############################################

DFT calculations were performed to calculate the density of states (DOS) (see Fig.~S3 in the Supplementary Information). The valence band (VB) of all \ce{Ga2O3} polymorphs is dominated by O $2p$ states, with small contributions from Ga $3d$ and Ga $4s$ states at the top and bottom of the VB. This is in agreement with previous calculations for some of the polymorphs \cite{FurtmullerPRB2016} and similar to other conventional oxide semiconductors \cite{MuddPRB2014,VasheghaniPRB2014,KingPRB2009}. In order to investigate the structure of the occupied states experimentally, VB spectra were collected using both SXPS and HAXPES. Fig. \ref{softhardcomp}a-b show the calculated and experimental VB regions for the $\alpha$, $\beta$, and $\varepsilon$ polymorphs, respectively. Fig. \ref{softhardcomp}a includes crosssection-weighted partial (pDOS) and total (tDOS) density of states for both SXPS and HAXPES. To aid direct comparison of theory and experiment, the DFT results were corrected for photoionisation cross sections and broadened to account for experimental resolution (see Supplementary Information for details). Both theory and experiment are aligned to the VB maximum VBM at 0 eV. The differences in cross sections between SXPS and HAXPES help to identify specific features and orbital contributions to the overall DOS. When going from soft to hard X-ray excitation energies, states are discriminated, e.g. in HAXPES the Ga $4s$ contributions are enhanced, validating DOS calculations. The sharper onset of the VBM predicted by theory for the $\alpha$-phase is clearly visible in the SXPS experimental results. In addition, the gradient of the VBM leading edge of the experimental data becomes progressively less steep when moving down the series from the $\alpha$ to $\beta$ to $\epsilon$ phases, following the trends observed in the theory and presenting a direct result of the variations in Ga coordination environments. This has obvious implications for determining the VBM to Fermi level separation experimentally, where often a linear extrapolation to the leading VBM edge is performed, severely underestimating the true value as was shown previously for $\beta$-\ce{Ga2O3} \cite{SwallowAPL2019}. Due to the severe drop in O $2p$ and Ga $3d$ cross sections which dominate the VBM, this is less pronounced in the HAXPES data. The features in the middle of the valence region are well represented in the theoretical calculations. $\alpha$-\ce{Ga2O3} has a distinct feature just below 4 eV, which is particularly clear in the SXPS spectra. This feature is deeper into the spectrum than the two seen in $\beta$-\ce{Ga2O3} at $\sim$2 eV and $\sim$3.5 eV, and the single feature in $\varepsilon$-\ce{Ga2O3} at $\sim$2 eV. Finally, the feature at the bottom of the VB, which is dominated by Ga $4s$ states, as clearly highlighted by the cross section differences, aligns well with the theoretical energy positions and overall shape predicted. Beyond aiding clear identification of the energy positions and characters of the contributions to the valence states, HAXPES data is less influenced by surface states and is a better representation of the bulk of the materials.\\

In order to highlight the effect of structural differences in the polymorphs on their respective electronic structure, the atomically resolved DOS for octahedrally and tetrahedrally coordinated Ga sites in $\beta$-\ce{Ga2O3} was calculated (see Fig.~\ref{softhardcomp}c). The main differences observed are located in the middle region of the VB, where two clear features at $\sim$2 eV and $\sim$4 eV appear. The octahedral DOS has much lower overall intensity in this region, which is attributed to the different distribution of Ga $d$ levels. Whilst similar differences are seen for the Ga $p$ states in the non cross-section corrected DOS (see Fig.~S4 in the Supporting Information), the $d$ states are greatly amplified due to cross section effects, resulting in much higher intensity in the middle of the VB under tetrahedral Ga coordination. Combination of the two calculated spectra using appropriate weightings of octahedral and tetrahedral contributions allows the simulation of spectra for the different polymorphs of \ce{Ga2O3}. It is important to note that due to bond length and bond angle variations across the polymorphs, this approach does not yield perfect representations of the polymorphs other than the $\beta$-phase, but it shows instructively the trend observed experimentally where the intensity diminishes in the middle of the VB as more octahedral spectral weight is added to the $\varepsilon$ and $\alpha$ phases, which have 75\% and 100\% octahedrally-coordinated Ga, respectively. This further confirms the influence of the local coordination geometry on the electronic structure observed.\\

\begin{figure*}[hbt!]
\centering
\includegraphics[width=0.98\textwidth]{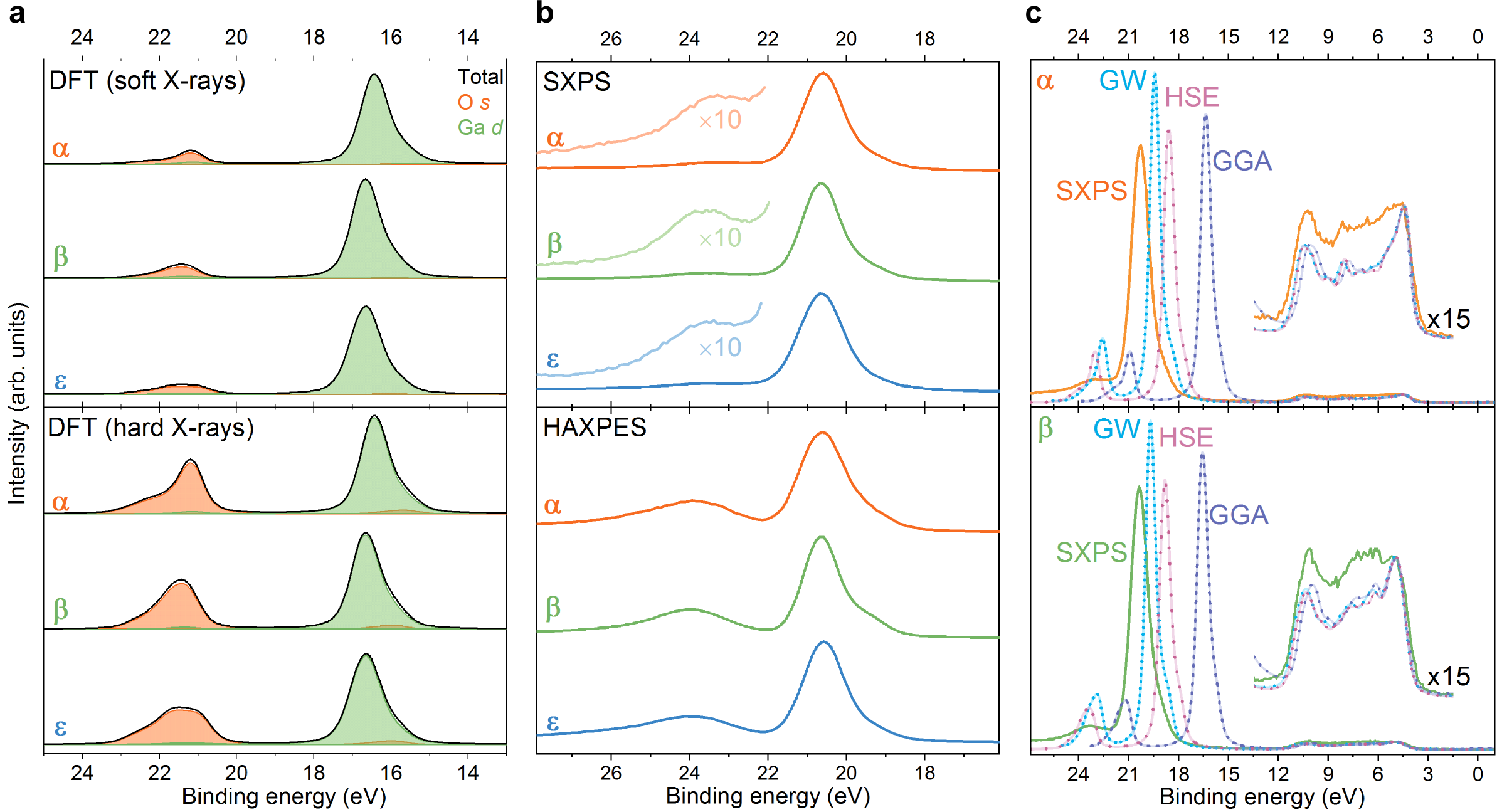}
\caption{\textbf{Semi-core levels of the \ce{Ga2O3} polymorphs. a,} Calculated pDOS and tDOS corrected for the photoionisation cross sections of soft and hard X-ray photoelectron spectroscopy measurements. \textbf{b,} SXPS and HAXPES data. The SXPS inset shows magnified (x10) views of the O $2s$ semi-core state. \textbf{c,} Comparison of DFT-GGA, HSE06 with 32\% exact exchange, and quasiparticle GW calculated tDOS with SXPS data for $\alpha$ and $\beta$-\ce{Ga2O3}. The inset shows an expanded view of the VB region. All calculated spectra are aligned to the leading edge of the experimental VB spectra.}
\label{ga3dpeak}
\end{figure*}

%############################################## 892 words
\subsection{Semi-Core States}
%##############################################

Semi- or shallow core states derived from $d$-orbital character in metal oxide semiconductors can have a large influence on the structure of the valence band itself \cite{Weiprb1988} and are therefore an important contributor to the final electronic structure. For \ce{Ga2O3}, the calculated DOS (see Fig.~S3 in the Supplementary Information) show that the semi-core levels consist of two peaks of O $2s$ and Ga $3d$ character. Fig.~\ref{ga3dpeak}a and b show the broadened and cross-section corrected DOS from DFT-GGA (Generalized Gradient Approximation) as well as the SXPS and HAXPES data for the semi-core level regions of the three \ce{Ga2O3} polymorphs. The calculated DOS have been aligned to the VBM of their respective polymorph. As with the valence spectra above, the combination of SXPS and HAXPES can help to identify specific contributions and orbital character to verify theoretical predictions.\\

The narrow and intense semi-core peaks in the DOS originate from a large number of dispersionless bands. In SXPS, the spectra of the polymorphs appear very similar. The largest difference between the set of calculated DOS is the shape of the O $2s$ derived peak at $\sim$21.5 eV. As previously mentioned, the $\varepsilon$-phase has the highest density of bands due to the larger number of O and Ga inequivalent sites, which explains why in Fig. \ref{ga3dpeak}a the O $2s$ peak is comparatively broader than those in the other two polymorphs. In Fig. \ref{ga3dpeak}b the HAXPES spectra show a much stronger contribution measured from the O $2s$ states than the SXPS counterparts do, due to the slower decrease in photoionisation cross section of O $2s$ relative to Ga $3d$, confirming the orbital nature of the states. Still, the O $2s$ peak is broad and relatively featureless compared to theoretical results. To retain consistency the same broadening as for the valence states was applied; however, the semi-core levels are more deeply bound and have a higher lifetime broadening relative to the valence states, resulting in the overall line widths of the spectral features observed \cite{Leyprb1974}. The difference in line width is much more subtle for the Ga $3d$ peak at $\sim$16.5 eV in the DFT-simulated SXPS in Fig.~\ref{ga3dpeak}a (or 20.7 eV in the SXPS in Fig. \ref{ga3dpeak}b). A small shoulder on the lower binding energy (BE) side of the Ga $3d$ peak in the experiment (see Fig.~\ref{ga3dpeak}b) may be misinterpreted as a second oxidation state of Ga, but in fact arises from $s$-$d$ hybridisation \cite{FurtmullerPRB2016}, much in the same way as seen in \ce{In2O3} \cite{KingPRB2009,FuchsPRB2008}.\\

Comparing theory and experiment, the O $2s$ peak is somewhat overestimated in intensity relative to the Ga $3d$ peak. This may be explained by an underestimation of the level of hybridisation between the two orbitals, as suggested previously for \ce{Ga2O3} and other oxides \cite{Navarroprb2015,KingPRB2009,FuchsPRB2008,shihprl2010}. This is further supported by the large disparity between the energetic positions of the two peaks, with the experimental data displaying $\sim$3 eV difference between the O $2s$ and Ga $3d$ states, while the theory predicts a difference around 5 eV. The energetic positions relative to the VBM of the semi-core levels are underestimated also. This is usually attributed to an overestimation of the $p$-$d$-repulsion, causing an underestimation of the band gap and consequently an underestimation of the position of these levels \cite{FuchsPRB2008,Fuchsprb2007}. Nevertheless, the shapes of the semi-core levels are well predicted by theory and clear evidence for $s$-$d$ hybridisation is demonstrated.\\

As the DFT-GGA calculations underestimate both the BE of the semi-core states and the $s$-$d$ hybridisation, quasiparticle (QP) electronic structure calculations within the GW approach were performed for a more detailed investigation \cite{HedinPR1965}. The semi-core states of $\alpha$- and $\beta$-phases are compared in Fig. \ref{ga3dpeak}c, but due to the structural complexity of the $\varepsilon$-phase, it was not possible to perform QP calculations. All spectra are aligned and normalised to the VBM of the experiment (composed of O $2p$ and Ga $3d$ character). As expected, the GW approach gives a better estimate of the absolute energy positions of semi-core levels compared to the DFT-GGA and HSE06 calculations for both polymorphs. A difference of $\sim3.8$ eV and $\sim3.9$ eV between the experimental and DFT-GGA calculated Ga $3d$ derived level for the $\alpha$ and $\beta$-phases, respectively, is reduced to $\sim$0.8 eV for both phases. This difference is consistent with other comparable calculations \cite{Fuchsprb2007,FuchsPRB2008,ShishkinPRB2007}, and the remaining underestimation has been suggested to be due to a small overestimation of the level of $p-d$ repulsion \cite{FuchsPRB2008}. It is worth noting that it could also relate to a (reduced) self-interaction error in the GW approximation \cite{Fuchsprb2007}, and may be offset by inclusion of vertex corrections \cite{FleszarPRB2005}. The GW approach also reduces the energetic difference between the Ga $3d$ to O $2s$ peaks to $\sim$3.2 eV and $\sim$3.3 eV for the $\alpha$- and $\beta$-phases, respectively, bringing them considerably closer to the experimental values of $\sim$2.9 eV and indicating the extent of hybridisation between these levels. The VB regions in both theory approaches are nominally the same with a very slight widening of the $GW$ spectra, which stems from a small ($<$ 0.3 eV) shift in the Ga $4d$ state at the bottom of the VB, in agreement with previous studies \cite{FuchsPRB2008,Fuchsprb2007,FurtmullerBook}. Overall the GW calculations give a more accurate description of the semi-core state energies whilst also retaining the intrinsic lineshape, but at higher computational cost. These results emphasise the importance of hybridisation in \ce{Ga2O3} and reinforce the effects that semi-core states have on the overall electronic structure.\\

\begin{figure}%[!ht]
\centering
\includegraphics[width=0.36\textwidth]{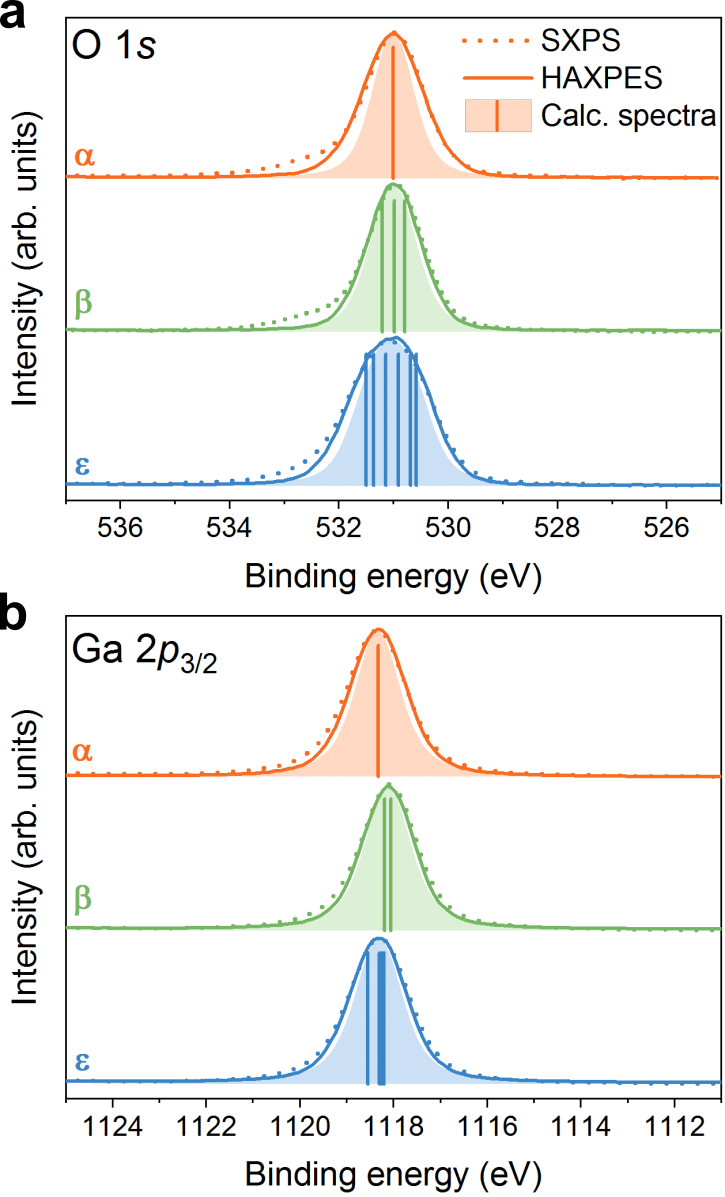}
\caption{\textbf{Core states of the \ce{Ga2O3} polymorphs. a,} O 1\textit{s} spectra. \textbf{b,} Ga 2\textit{p}$_{3/2}$ spectra. For each core state the SXPS (dotted line) and HAXPES (solid line) spectra as well as the calculated spectra (shaded) and calculated core level binding energies (vertical lines) for all atoms in the primitive unit cell of the different polymorphs are shown.}
\label{corelevels}
\end{figure}

%############################################## 820 words
\section{Core States}
%##############################################

An often overlooked source of information regarding the electronic structure of polymorphic materials is the core level spectra. Whilst large amounts of data regarding bonding and chemical shifts are available from sources such as Moulder \cite{Moulder1992}, it is often assumed the core levels will be identical across polymorphs as they include comparable chemical environments, e.g. Ga-O bonds in all \ce{Ga2O3} polymorphs. In theoretical investigations, usually the valence and possibly semi-core levels are investigated as they are the most heavily involved in determining the chemical and physical properties of a material, and are expected to be most sensitive to any structural differences. However, the line shapes of deeper core states can be explained on a fundamental level and strongly reflect the structural differences in polymorphs in parallel to the observations made in semi-core and valence states.\\

The O $1s$ and Ga $2p_{3/2}$ core levels from SXPS and HAXPES measurements of the three polymorphs are shown in Fig. \ref{corelevels} (survey spectra are shown in Fig.~S5 in the Supplementary Information). The O $1s$ spectra show a striking difference between the three polymorphs and the $\varepsilon$-phase has a greatly increased line width in both SXPS and HAXPES. The full width half maximum (FWHM) of the $\alpha$ and $\beta$ polymorphs is on average 1.2 eV compared to 1.6 eV for the $\varepsilon$ polymorph (see Fig.~S6 in the Supplementary Information for peak fits and FWHM values for all core levels). The influence of surface hydroxylation and other surface states, as well as any influence from surface band bending, can be excluded as the broadening is observed in both SXPS and HAXPES. Heavily doped metal oxide materials commonly display plasmon losses which can manifest as a shoulder to higher BE of the main core line \cite{BorgattiPRB2018} and this observed broadening of the $\varepsilon$-phase O $1s$ core level might be attributed to this initially. $\varepsilon$-\ce{Ga2O3} has a large band gap ($\sim$4.9 eV \cite{OshimaJAP2015}, similar to $\beta$-\ce{Ga2O3}), and can be extrinsically doped with e.g. Si or Sn resulting in carrier densities on the order of $3\times10^{18}$cm\textsuperscript{-3} \cite{Bardeleben19,Parsini19}. However, even degenerately doped \ce{Ga2O3} may not display plasmon features, as observed for Sn-doped $\beta$-\ce{Ga2O3} with a carrier density of $6\times10^{18}$cm\textsuperscript{-3} \cite{SwallowAPL2019}, above the Mott criterion (around $3\times10^{18}$cm\textsuperscript{-3}) where metallic behaviour should prevail. Since the samples in this study are not intentionally doped, the carrier densities should be below this limit, and hence no plasmon losses are expected. Furthermore, the $\varepsilon$-\ce{Ga2O3} Ga $2p_{3/2}$ core level is of similar width to that of the other two phases, further indicating the absence of plasmon losses. There are a very limited number of reports on the XPS of $\varepsilon$-\ce{Ga2O3} available for comparison \cite{MulazziAPL2019}. However, in this study several $\varepsilon$-\ce{Ga2O3} samples were grown with different MBE recipes. All analyzed films consist of columnar, orthorhombic rotational domains usually in the range of 2-10 nm for $\varepsilon$-layers deposited on c-plane sapphire \cite{CoraCEC2017}. The $\omega$-rocking curves of the investigated samples vary between 0.14$^\circ$ and 0.40$^\circ$ (Fig.~1f and Fig.~S2 in the Supplementary Information). Despite the related differences in the films microstructure, no considerable differences are found in the SXPS and HAXPES peak broadening among the samples.\\

The calculated Ga $2p_{3/2}$ and O $1s$ binding energies for all atoms in the unit cell and the corresponding spectra for each of the three polymorphs are presented in Fig.~\ref{corelevels} (see Fig.~S8 and related text in Supplementary Information for further details). The DFT-GGA approach is used here, leading to shifts in absolute energy positions, which are summarised in Tab.~S1 in the Supplementary Information, and the \textit{ab initio} calculations give core binding energies for O $1s$ and Ga $2p_{3/2}$ states for all inequivalent atomic sites. Therefore, the number of distinct core states observed is directly correlated to the number of inequivalent sites, which depends on the polymorphism. $\alpha$-\ce{Ga2O3} has only single Ga and O sites, with four symmetry related Ga lattice sites (Wyckoff position 4c) and six equivalent O sites (Wyckoff position 6e). All atoms in $\beta$-\ce{Ga2O3} occupy Wyckoff positions 4i, yielding two inequivalent Ga and three inequivalent O positions. The energies of the inequivalent positions are closely spaced with maximum energy differences $\Delta E_{Ga}=0.137$ eV and $\Delta E_{O}=0.413$ eV. Finally, the primitive unit cell of $\varepsilon$-\ce{Ga2O3} has 16 Ga atoms with four different Wyckoff positions 4a, and 24 O atoms at six different Wyckoff positions 4a. This results in four inequivalent energy levels for Ga $2p_{3/2}$ and six energy levels for O $1s$ with $\Delta E_{Ga}=0.346$ eV and $\Delta E_{O}=0.917$ eV. These differences in core state binding energies caused by the local coordination environments in \ce{Ga2O3} explain the broadening observed experimentally, particularly in the O $1s$ spectra of the $\varepsilon$-phase.\\

While the effect of polymorphism on the density of states in the valence and semi-core region has been reported previously \cite{KingPRB2009,FurtmullerPRB2016,FuchsPRB2008,YoshiokaJP2007}, such an effect on the core spectra has rarely been observed. The combination of high quality samples, SXPS and HAXPES to exclude surface effects, and the complementary theory approach enable a deeper understanding of the direct relationship between local coordination geometry and spectral features. These differences are obvious in the core levels and complement valence band data, which has contributions from many orbitals. This shows that core states can act as a direct probe for differences in local structure in polymorphic systems.\\

%############################################## 560 words
\section{Unoccupied States}
%##############################################

It is clear from the discussion so far that distinct differences in spectral shape between the three polymorphs can be evidenced in the valence, semi-core and core state spectra and are directly related to the local coordination environments. Beyond the occupied states, it is essential to probe unoccupied states in the conduction band (CB) of the polymorphs to determine further differences in the electronic structure brought about by the varying crystal structures. The CB states are experimentally probed using X-ray absorption spectroscopy (XAS), which is complimentary to XPS. While XPS spectra yield information on the VB structure and can be calculated within DFT, the XAS final state is a neutral core excited state. As such, the BSE approach is used here, which explicitly includes the effects of the electron-core hole interaction in the many-electron system \cite{Vorwerk2017,Vorwerk2019}. To explore the advantages of BSE over lower cost DFT + core-hole correction, both methods were initially employed. Whilst both give good descriptions of the spectra, a clear improvement of agreement between theory and experiment is found with BSE (see Fig.~S10 in Supplementary Information). The calculations show that the effect of the electron-hole interaction on the spectra is strong, as the intense peak at the absorption onset is missing when electron-hole correlation is neglected.\\

\begin{figure}%[ht]
\centering
\includegraphics[width=0.36\textwidth]{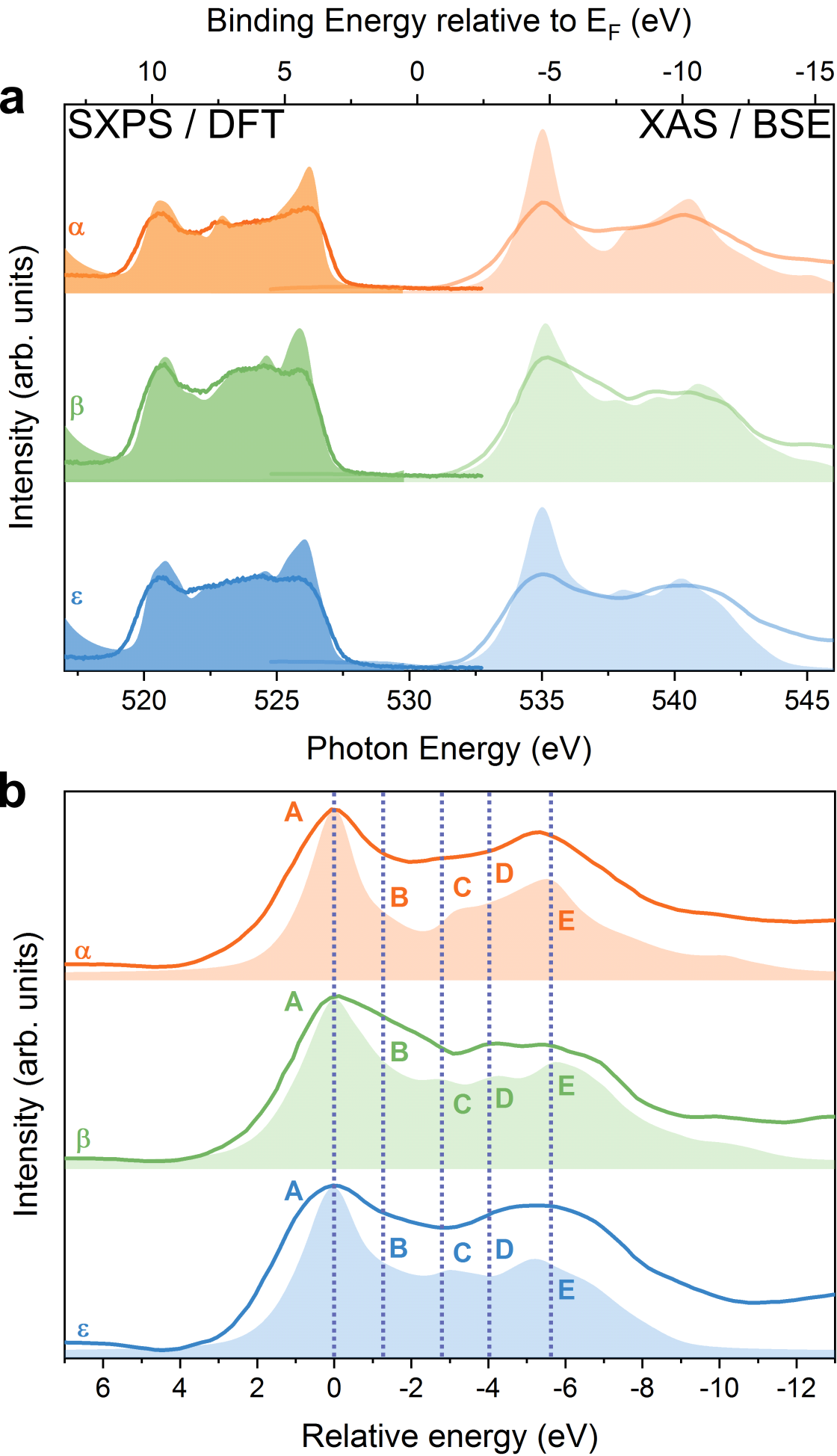}
\caption{\textbf{Conduction and valence band states of the \ce{Ga2O3} polymorphs. a,} SXPS and DFT of the VB as well as O XAS and BSE of the CB plotted on a common photon energy axis. Theory and experiment were normalised to their respective areas. The Fermi energy $E_F$ position as determined from SXPS is included and the top x axis gives binding energy relative to the $E-F$ position. \textbf{b,} XAS O K-edge spectra (lines) are combined with excited state O K-edge absorption spectra from BSE (shaded area). Data are displayed on a relative energy axis where feature A is aligned at 0 eV. For both plots experimental data is shown as lines, theoretical data as shaded areas. }
\label{XAS}
\end{figure}

Fig.~\ref{XAS}a shows the XPS and XAS experimental results together with the respective theoretical results from DFT and BSE on a common photon energy scale (see Fig.~S11 and related text in the Supplementary Information on details of the alignment). The Fermi energy $E_F$ position as determined from the SXPS measurements is plotted, and appears within the band gap towards the CBM (consistent with reports for $\beta$-\ce{Ga2O3} from particle irradiation studies \cite{Vines2018}), indicating all three materials are n-type, but not degenerate. This diverges from what is expected for other common metal oxide systems, which often display Fermi level degeneracy\cite{King2010}. Whilst the combination of XPS and XAS cannot be used to determine absolute band gap values due to the uncertainties in determining the energy alignment and positions, it does allow a qualitative discussion of the electronic structure. Overall, subtle differences in the separation of the VBM and CBM are observed. The VBM and CBM onsets can be determined from second order derivatives of the spectra (see Fig.~S12 in the Supplementary Information) with some uncertainty stemming from even minimal noise levels in the data as well as instrumental broadening. This method yields VBM-CBM energy separations around 5 eV for the three polymorphs depending on the choice of onset position, which although not perfectly representative for the band gaps lies close to their reported values for \ce{Ga2O3} \cite{OshimaJAP2015,ROBERTSJCG2018,ZhangAPL2016}. This alignment procedure does not take into account the interaction of the excited electron and its core hole, which can affect the positioning of the spectra, typically lowering the XAS spectra onset up to 0.5-1.0 eV \cite{Zhang12,SallisAPL2013}. However, in the present case the energy gaps observed are more consistent with the band gap indicating a smaller core-hole effect as expected for a wide band gap weakly correlated oxide.\\ 

In order to discuss the O K-edge absorption spectra, which are derived from 1\textit{s}$\rightarrow$2\textit{p} transitions, in more detail, the experimental results are combined with calculated XAS spectra in Fig.~\ref{XAS}b. Data are aligned relative to the most intense feature A (initial XAS calculations can be found in Fig.~S9 in the Supplementary Information). Much like in the VB spectra, the $\beta$ and $\varepsilon$-phases share similar O K-edge shapes, whilst the $\alpha$-phase is distinctly different in its features. The total O K-edge spectra are the sum of the absorption spectra of the inequivalent oxygen atoms in a given polymorph and all spectra show five identifiable features A-E. The first-principle calculations used here allow the identification of atomic contributions of the inequivalent oxygen atoms to the different features in the spectrum. Feature A, which occurs in all three polymorphs, is due to transitions from the O $1s$ states to low-energy conduction bands and occurs for each inequivalent atom. Here, strong interactions between the excited electron and the core hole are found, which yield the sharp intense feature that is not found in the projected DOS (see Fig.~S3 in the Supplementary Information). Features B and C in the spectra of the $\beta-$ and $\varepsilon$-phases originate from transitions from the $1s$ states of oxygen atoms that are predominately bonded to tetrahedrally coordinated Ga atoms. In the $\alpha$-phase, this coordination does not occur and peak B at $\sim$1.2 eV above A is missing, while peak C has greatly reduced in intensity. Features D and E, on the other hand, originate from transitions at oxygen atoms that are predominately bonded to octahedral Ga atoms. Consequently, these features have the strongest relative intensity for the $\alpha$-phase. Overall, the features are well reproduced by the calculations in both their intensities and energy positions and the observed differences confirm a significant influence of the local coordination environments on the occupied states, both directly at the CBM and beyond.\\

%############################################## 260 words
\section{Conclusion}
%##############################################

This work presents a systematic, in-depth spectroscopic and theoretical investigation of the structure - electronic structure relationship in three polymorphs of \ce{Ga2O3}: the $\alpha$, $\beta$, and $\varepsilon$-phases. State-of-the-art, multi-level theoretical approaches together with advanced X-ray photoelectron and absorption spectroscopy methods provide the first combined theory-experiment investigation of a range of \ce{Ga2O3} polymorphs.\\

Based on purely stoichiometric arguments one might expect the electronic structures and core states to be the same for different polymorphs, however, the present work clearly shows that valence, semi-core and core states are all influenced by differences in local coordination environments. The variations in octahedrally and tetrahedrally coordinated Ga centres in the polymorphs result in discernible changes in the density of states for both valence and conduction band states. Particularly for semi-core and core states the importance of higher level theoretical approaches to rationalise experimentally observed features is demonstrated. Throughout, the $\alpha$-phase proves to be intrinsically different from the $\beta$ and $\varepsilon$ phases due to it only having octahedrally coordinated Ga sites. The combination of SXPS and HAXPES enables the exclusion of pure surface effects further highlighting that observed spectral features are directly connected to the bulk crystal structure.\\

The detailed, fundamental information on the \ce{Ga2O3} polymorphs presented here provides a solid basis for advanced studies of \ce{Ga2O3} and related oxide materials and lays important groundwork for further optimisation and implementation of \ce{Ga2O3} polymorphs in device architectures. It is clear that the choice of polymorph and the optimisation of the local coordination environments can be used to tune the electronic structure of \ce{Ga2O3}, essential for its ultimate behaviour within devices.\\

%##############################################
\section{Methods}
%##############################################

%##############################################
\subsection{Growth and Structure}
%##############################################

Various methods were used to produce optimum, high quality samples for the wide range of \ce{Ga2O3} structures discussed in this paper. The conductivity of the samples was of special concern to allow photoelectron spectroscopy measurements. Bulk $\beta$-\ce{Ga2O3} single crystals in the (010) orientation were obtained from Novel Crystal Technology Inc., Tamura Corporation, and were grown from the melt using the edge-defined film-fed growth method \cite{AidaJSAP2008}. The bulk crystals are unintentionally-doped and lightly Sn-doped samples are used as a reference to ensure sufficient conductivity in order to avoid sample charging effects during photoelectron measurements. The spectra were essentially identical indicating charging was not an issue for the bulk crystals. The net donor density ($N^+_D$-$N_A^-$) was determined by capacitance-voltage measurements to be $\sim$10\textsuperscript{18} cm\textsuperscript{-3} in each. $\alpha$-\ce{Ga2O3} was deposited via plasma enhanced atomic layer deposition (PE-ALD) onto c-plane sapphire substrates, displaying mainly (0001) orientation, at a constant temperature of 300$^{\circ}$C giving a final film thickness of 30 nm as determined by spectroscopic ellipsometry and cross-sectional transmission electron microscopy \cite{ROBERTSJCG2018,ROBERTS2019125254}. $\varepsilon$-\ce{Ga2O3} was grown via metal-exchange catalyzed (MEXCAT)-MBE mediated by either indium \cite{Vogt2017} or tin \cite{KrachtPRA2017}, on c-plane sapphire substrates. The thicknesses of the indium mediated films was measured by in-situ laser reflectometry during growth and values of 100 nm ($\varepsilon_1$ in Supplementary Information) and 800 nm ($\varepsilon$ samples included in the main manuscript and labelled as $\varepsilon_2$ in the Supplementary Information), respectively, were found. The tin mediated thin film ($\varepsilon_3$ in Supplementary Information) has a thickness of 800 nm as determined by optical reflectometry. XRD and core level XPS data from all samples are included in Figs.~S2 and S7 in the Supplementary Information.\\

Phase and crystallinity of the samples were studied under ambient conditions by \textit{ex-situ} X-ray diffraction of symmetric out-of-plane reflections using monochromated X-rays (Cu K$_{\alpha1}$) in parallel-beam geometry. Out-of-plane $\theta$:2$\theta$ scans between 10$^{\circ}$ and 90$^{\circ}$ were performed to identify the \ce{Ga2O3} phase and orientation, and  rule out the existence of secondary phases. A $\omega$ rocking curve centred about the \ce{Ga2O3} peak was carried out to assess the degree of the \ce{Ga2O3} crystallinity. The $\beta$ and $\varepsilon$-\ce{Ga2O3} samples were measured in a four-circle lab diffractometer (PANalytical X'Pert Pro MRD) using a 1 mm wide detector slit. A Rigaku SmartLab X-ray diffractometer with a 9 kW rotating anode was used for $\alpha$-\ce{Ga2O3}. Higher-resolution $\theta$:2$\theta$ between 37$^{\circ}$ and 43$^{\circ}$ were measured to assess film thickness of 30 nm, from the Pendell\"{o}sung fringes about the strong $\alpha$-\ce{Al2O3} 0006 substrate peak and the main $\alpha$-\ce{Ga2O3} 0006 peak.\\

%##############################################
\subsection{Electron and X-ray Spectroscopy}
%##############################################
Laboratory-based soft X-ray photoelectron spectroscopy (SXPS) was performed on a Thermo Scientific K-Alpha+ X-ray photoelectron spectrometer (base pressure 2 $\times$ 10\textsuperscript{-9} mbar), utilising a monochromated microfocused Al K$\alpha$ X-ray source (h$\nu$ =1486.7 eV) with a spot size of 400 $\mu$m and a 180$^\circ$ double focusing hemispherical analyzer with 2D detector. The X-ray source was operated at 6 mA emission current and 12-kV anode bias. Pass energies of 20 eV for core level and 15 eV for valence band spectra were used. HAXPES data were collected at the I09 beamline at Diamond Light Source, UK \cite{Lee2018}. A photon energy of 5.93 keV was used for all measurements. A double-crystal Si (111) monochromator was combined with a Si (004) channel-cut crystal as a post-monochromator to achieve the final energy resolution. The end station is equipped with a VG Scienta EW4000 electron analyzer with a wide acceptance angle of $\pm$28$^\circ$ and a pass energy of 200 eV was used. All measurements were performed in grazing incidence geometry at angles below 5$^\circ$ between the incoming X-ray beam and the sample surface. The resolution of SXPS and HAXPES measurements as determined by the width of the Fermi edge of gold are 0.44 eV and 0.25 eV, respectively. X-ray absorption spectroscopy (XAS) of the O K-edge in total electron yield mode (TEY) was performed on the same beamline as the HAXPES measurements, without the requirement of a secondary detector.\\

%##############################################
\subsection{Computational Details}
%##############################################
The \textit{ab initio} calculations are performed with two different computer packages. The Kohn-Sham band structure of $\alpha$-, $\beta$-, and $\varepsilon$-\ce{Ga2O3} is obtained from all-electron full-potential DFT calculations using linearized augmented plane-wave (LAPW+LO) basis functions as implemented in the \texttt{exciting} code \cite{GulansJPCM2014}. Here, the basis set size is described by the dimensionless parameter $R_{MT}|\mathbf{G}+\mathbf{k}|_{max}$, set to 8 for $\alpha$- and $\beta$- and to 10 for $\epsilon$-\ce{Ga2O3}. For the $\alpha$- and $\beta$-phase, the calculations are performed on an $8\times 8 \times 8$ $\mathbf{k}$-mesh in reciprocal space. Since the primitive unit cell of $\varepsilon$-\ce{Ga2O3} is larger than those of the other two, a $4 \times 4 \times 4$ $\mathbf{k}$-grid is used. The generalized gradient approximation to the exchange-correlation functional as developed by Perdew, Burke, and Ernzerhof is employed \cite{PBEsol2008}. All calculations are performed using experimental lattice parameters \cite{Lipinska08}.

The X-ray absorption spectra at the O K edge are calculated within many-body perturbation theory by solving the Bethe-Salpeter equation (BSE) \cite{Vorwerk2017,Vorwerk2019} which explicitely includes the screened interaction between the electron  and the core hole as well as local-field effects (LFE). For the $\alpha$- ($\beta$-) phase, a shifted $9 \times 9 \times 9$ ($10 \times 10 \times 10$) $\mathbf{k}$-grid is used, and LFE are included up to a cut-off $|\mathbf{G}+\mathbf{q}|_{max}=3 a_0^{-1}$ (1 $a_0^{-1}$), where $a_0$ is the Bohr radius. For  both phases, the 20 lowest conduction states are included in the construction of the BSE Hamiltonian. For the $\epsilon$-phase, the calculations are performed on a shifted $8 \times 8 \times 8$ $\mathbf{k}$-grid using the lowest 60 unoccupied states in the BSE Hamiltonian. A cut-off of $|\mathbf{G}+\mathbf{q}|_{max}=1.5 a_0^{-1}$ is employed to include LFE.  For the XAS calculations of all three polymorphs, the screening in the random-phase approximations is computed using 100 unoccupied bands. A Lorentzian broadening of 0.15 eV is applied to the final spectra to mimic the lifetime broadening of the excitations. For the $\beta$- and $\epsilon$-phase, a spectrum is computed for each inequivalent oxygen atom, and the final spectra are obtained as the sum of the atomic contributions. The BSE results are broadened with a Gaussian-Lorentzian convolution, and the onsets are  shifted so that the energy position of the respective leading edge matches that of the experiment. The computational data obtained with the \texttt{exciting} code is available in the NOMAD Repository~\cite{NOMADalpha, NOMADbeta, NOMADepsilon}. In Figs.~\ref{corelevels} and \ref{XAS}, the binding energies are aligned to the SXPS O 1\textit{s} core level; a 1 eV shift is applied to the XAS in Fig.~\ref{XAS} to account for core-hole effects. The corresponding photon energies are plotted along the top axis.

To obtain the quasiparticle (QP) band structure for the $\alpha$- and $\beta$-polymorphs in Fig.~\ref{ga3dpeak}, Hedin’s GW approach is employed. \cite{HedinPR1965}. Single-shot $G_0W_0$ calculations are performed within the frozen-core projector-augmented wave (PAW) approach as implemented in the Vienna ab initio simulation package (VASP) \cite{KressePRB1996,KressePRB1999}. The QP calculations use DFT calculations with the HSE06 hybrid functional \cite{heydJCP2003, Heyd2006} as a starting point since HSE06 provides better starting wave functions for the for the electronic structure. In these calculations, the reciprocal screening length of the HSE06 functional is altered from the original value 0.2~\AA$^{-1}$ to 0.3~\AA$^{-1}$ as proposed by Fuchs \textit{et al.}~\cite{Fuchsprb2007}. The Ga~$4s$, Ga~$4p$ states, and the Ga~$3d$ core states are treated as valence states. A plane-wave cutoff of 410~eV is employed together with a $\Gamma$-shifted $8\times 8\times 8$ {\bf k}-grid and a Gaussian broadening with a smearing width of 0.1~eV. For the G$_0$W$_0$ calculations 512 bands, i.e. 44 occupied and 468 unoccupied bands, are included in order to converge the dielectric function. The full frequency-dependent and wave-vector-dependent screening is used without any approximations. It is sampled on a (non-uniform) frequency grid of 96 energies, ranging from 0 to 192 eV, and a plane-wave cutoff of 150~eV was employed to describe its wave-vector dependence. A full explanation of the methodology used here to evaluate the QP electronic structure is given elsewhere \cite{FurtmullerPRB2016}. The underlying geometries are optimized using DFT-GGA with the AM05 exchange correlation functional \cite{ArmientoPRB2005} and a $11\times 11\times 11$ {\bf k}-grid. The resulting parameters agree closely with experimental ones. To construct the DOS for the octahedrally and tetrahedrally coordinated gallium atoms, the atomically-resolved DOS was calculated for the primitive unit cell of $\beta$-Ga$_2$O$_3$ as a template structure, using the HSE06 hybrid functional \cite{heydJCP2003, Heyd2006}. For the stand-alone HSE06 calculations in the site-projected DOS and included in Fig. \ref{ga3dpeak}(c), the fraction of exact-exchange was increased from 25\% to 32\% to better reproduce the experimental band gap and lattice constants, as reported and described in Ref.~\cite{Peelaers18}.\\

%##############################################
\section{Acknowledgements}
%##############################################
J.E.N.S.  acknowledges  funding  through  the  Engineering and  Physical Sciences  Research  Council  (EPSRC)  Centre  for Doctoral Training in New and Sustainable Photovoltaics (EP/L01551X/1). AR acknowledges the support from the Analytical Chemistry Trust Fund for her CAMS-UK Fellowship. We acknowledge Diamond Light Source for time on Beamline I09 under Proposals SI21430-1 and SI24670-1. The authors would like to thank Dave McCue, I09 beamline technician, for his support of the experiments. S.A.R. and L.F.J.P. acknowledge support from the National Science Foundation DMREF-1627583. Z.W.L.-H. gratefully acknowledges support from a Doctoral Fellowship in Residence Award from the ALS. XES and XAS were performed at beamline 8.0.1 at the Advanced Light Source. This research used resources of the Advanced Photon Source, a U.S. Department of Energy (DOE) Office of Science User Facility operated for the DOE Office of Science by Argonne National Laboratory under Contract No. DE-AC02-06CH11357. The Liverpool XRD facility used was supported by the EPSRC (EP/P001513/1). The authors would like to thank M. Kracht for assistance with Sn-mediated MBE growth of the $\epsilon$-\ce{Ga2O3} sample. Work by C.D., C.V., O.B., P.M., and P.V. was performed in the framework of GraFOx, a Leibniz-ScienceCampus partially funded by the Leibniz association. The work of J.B.V. was performed under the auspices of the U.S. DOE by Lawrence Livermore National Laboratory under contract DE-AC52-07NA27344. 

%##############################################
\section{Supporting Information}
%##############################################
Detailed diffraction patterns, computational details, corrections and results of calculated valence DOS, SXPS and HAXPES survey spectra, fitted O 1\textit{s} and Ga 2\textit{p}$_{3/2}$ core level spectra, O 1\textit{s} and Ga 2\textit{p}$_{3/2}$ core level spectra of all $\varepsilon$ samples, computational and correction details for core level calculations, computational details for XAS calculations and comparison of results from DFT core-hole and BSE approach, and alignment of XAS and XPS data using XES

%##############################################
\section{Author Contributions}
%##############################################
A.R. conceived the experiment. P.M., P.V., O.B., A.K., M.E., J.S., M.R.W., J.W.R., P.R.C., M.J.S. and P.M. synthesized, performed characterisation and selected the appropriate \ce{Ga2O3} samples. P.M., T.D.V., P.V., M.J.S. and P.M. conducted the XRD experiments. A.R. conducted the SXPS experiment. J.E.N.S., P.K.T., T.-L.L., L.A.H.J. and A.R. conducted the HAXPES/XAS experiments at Diamond Light Source. J.E.N.S., O.B. and A.R. analyzed the experimental data with input from T.D.V.. C.V., J.F., J.B.V., S.A.R., L.F.J.P., and C.D. performed the {\it{ab initio}} calculations.  J.E.N.S., C.V., O.B. and A.R. wrote the manuscript with input from all authors.

%##############################################
\section{Competing Interests}
%##############################################
The authors declare no competing interests.

%apssamp
%##############################################
\bibliographystyle{naturemag}
\bibliography{ga2o3}% Produces the bibliography via BibTeX.
%##############################################

\newpage
\textbf{Table of Contents (ToC Image)}\\

\begin{figure}%[ht]
\centering
\includegraphics[width=8.47cm]{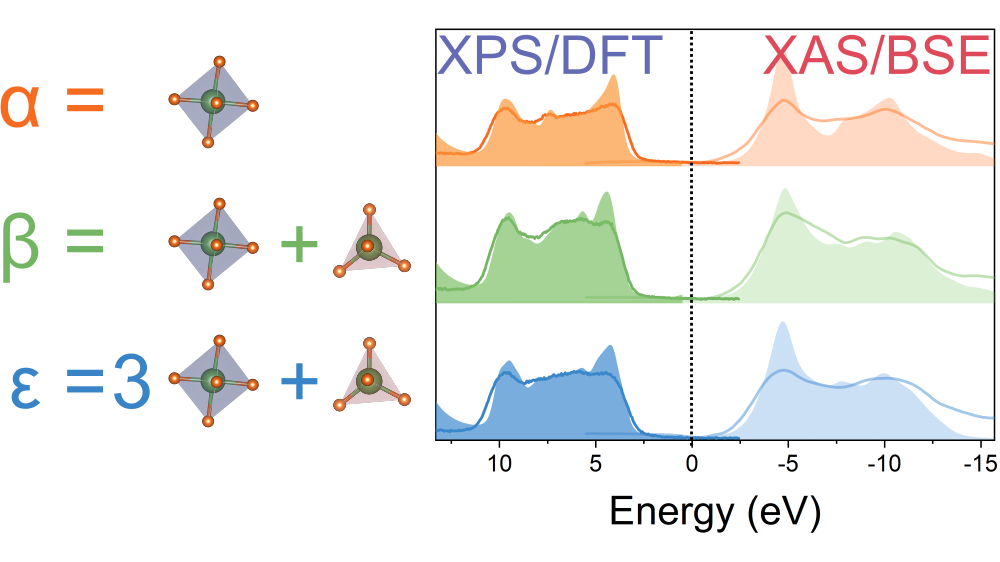}
\end{figure}

\end{document}